# Sources of experimental errors in the observation of nanoscale magnetism.

M. A. Garcia[1,2 a)], E. Fernandez Pinel[1,3], J. de la Venta[1,3], A. Quesada[1,4], V. Bouzas[1], J. F. Fernández[2], J. J. Romero[2], M. S. Martín González[5], J. L. Costa-Krämer[5].

[1] Dpto Física de Materiales, Universidad Complutense de Madrid, Madrid, Spain.
[2] Instituto de Cerámica y Vidrio, CSIC, Madrid, Spain.
[3] Instituto de Magnetismo Aplicado, Universidad Complutense de Madrid, Madrid, Spain.
[4] Instituto de Ciencia de Materiales de Madrid, CSIC, Madrid, Spain.
[5] Instituto de Microelectrónica de Madrid, CSIC, Tres Cantos, Madrid, Spain.

[a)] Electronic mail: ma.garcia@fis.ucm.es

## Abstract

It has been recently reported that some non-magnetic materials in bulk state, exhibit magnetic behavior at the nanoscale due to surface and size effects. The experimental observation of these effects is based on the measurement of very small magnetic signals. Thus, some spurious effects that are not critical for bulk materials with large magnetic signals may become important when measuring small signals (typically below $10^{-4}$ emu). Here, we summarize some sources of these small magnetic signals that should be considered when studying this new nanomagnetism.

## 1. Introduction

It is well known that the physical properties of the materials are modified when the size is reduced to the nanoscale due to size and surface effects. In the last years, with the development of new techniques to fabricate, manipulate, measure and image nano-objects, a large number of experiments have shown new optical, electrical, magnetic and mechanical properties for these nano-objects. Probably, the most remarkable modifications at the nanoscale appear in the magnetic properties: besides the quantitative modifications of the magnetic properties of the nanomaterials, it has been claimed that many materials show magnetic behavior at the nanoscale despite their bulk non-magnetic character [1,2,3,4,5,6,7,8,9,10,11,12, 13,14], including for instance all kind of oxides [15] and superconductors [16]. Some authors even suggested that this nanoscale magnetism may appear in any kind of material [16]. While some of these experiments are easily reproducible and have been reported by different groups independently, many of the new findings are hard to reproduce, creating controversy and leading to a confusing picture. This is particularly true in the case of magnetism in oxides and semiconductors, where erroneous experimental results have produced a considerable puzzlement. Identification of the right and reproducible new effects and their separation from flawed science [17], is crucial to advance in the discovery and applications of this nanoscale magnetism.

Many incorrect results arise from the experimental difficulties to measure these magnetic moments at the nanoscale. Classical magnetic materials (3d elements and rare earths) exhibit magnetic moments of the order of a Bohr magneton ($\mu_B$) per atom; thus, even for 1 mg, the magnetic moments of the samples are usually over $10^{-2}$ emu. On the contrary, most of the experiments on the new magnetism report very low magnetic moments of the order of $10^{-2}$-$10^{-3}$ $\mu_B$ per atom, two or three orders of magnitude smaller than traditional bulk ferromagnets. As synthesis methods of nanostructures hardly produce more than few milligrams, the magnetic moments to be experimentally detected are about $10^{-4}$-$10^{-6}$ emu. The situation becomes even worse for magnetism of surfaces and interfaces: In a 3mm x 3mm surface (typical area of magnetometers sampleholders) the number of atoms is ~ $10^{14}$, with a mass of the order of one microgram so a signal of ~$10^{-6}$ emu would correspond to giant magnetic moments per surface atom. In this situation, it is necessary to use the best experimental setups working at the limit of its resolution to detect those new magnetic moments. The presence of small measuring artifacts, not so important when measuring traditional bulk ferromagnets, can modify completely the results leading to surprising but incorrect new findings: a very small false magnetic signal will become huge when normalized to the sample mass. X-ray magnetic circular dichroism (XMCD) [18] could be a solution for this problem due to its element specificity and sensitivity to detect the magnetization from a few number of atoms. Neutron scattering offers also outstanding possibilities for the study of this magnetism when available [19]. However, in many nano-objects only a small fraction of the atoms contributes to the magnetization of the sample, arising a dilution problem that renders XMCD unuseful to detect it (in addition to the fact that XMCD is available just at synchrotron facilities and it is not easy to get time for these measurements). In general, magneto-optical methods, can be very convenient in this research as they offer the possibility to select certain processes [20] by tuning the light wavelength or select different areas of the sample to detect inhomogeneous contamination [21]. Magnetotransport measurements can also help to confirm that the presence of magnetism is not due to impurities [13,14] but these experiments depends on the transport properties of the material and in principle can not be performed in any kind of material. Therefore, when the discovery of new magnetic effects at the nanoscale relays on conventional magnetometry measurements, the experimental work must be extremely careful.

We examine here several sources of small magnetic signals that can be disregarded for samples with large signals (over ~$10^{-3}$ emu) but must be considered when performing experiments with small ones, as it is usually the case of the nanoscale magnetism. We will focus just on signals arising from samples handling or measurements procedures but not on contamination due to starting materials purity, sample fabrication procedures nor systematic errors of measurement setups.

## 2. Experimental results

### 2.1 Kapton tape

Kapton® tape (KT) is commonly used to place and fix samples on sampleholders as it is removed cleanly leaving no residue and exhibits good thermal stability retaining its adherence at very low T (allowing to perform experiments scanning temperature in a wide interval) . KT consists on a polyimide film and adhesive system silicone or acrylic based, having nominally no magnetic atoms and expected diamagnetic behavior. In the following experiments, KT 1 cm wide has been used. This tape has a linear density of 9.58 mg/cm. Figure 1a shows the magnetization curves from several pieces of KT with length ranging between 3 cm and 9 cm, placed on a sampleholder and measured at 300 K.  The overall behavior is diamagnetic with a susceptibility of $4.3 \cdot 10^{-7}$ emu/g·Oe ($4.2 \cdot 10^{-9}$ emu/cm·Oe). The curves exhibit a superimposed ferromagnetic-like contribution (figure 1b) with a saturation magnetization ranging up to 5 $\mu$emu/cm as figure 1b illustrates. Measuring 20 pieces of KT shows that the FM signal does not scales accurately with the length of KT but it is  typically between 5 and 15 $\mu$emu (see figure 1c). This ferromagnetic (FM) contribution shows a weak dependence with the temperature being 20% larger at 5 K than at 300 K as shown in figure 1d.

In order to check the origin of this FM signal, a clean piece of KT that presented pure diamagnetic behavior was exposed to the air for 30 minutes and re-measured. After exposition, a ferromagnetic-like contribution of 10 $\mu$emu was found. Dust particles in the air contain typically 3.5% of iron atoms [22] which could account for this FM signal. A signal of $10^{-5}$ emu as the measured one could arise from a pure iron particle of ~15 μm. When the KT is stored in air we found the FM signal (above 2 $\mu$emu)  for 60-70% of the pieces, but if the tape is stored in a hermetic plastic bag since received from the supplier and only opened to be used, the frequency is reduced to 10%-20%. If the lab is a clean room (grade 100000) the frequency of this FM signal is also significantly reduced. Therefore, we may conclude that the FM signal observed in KT pieces of the order of 10 μemu is due to the adhesion of dust particles with iron content.

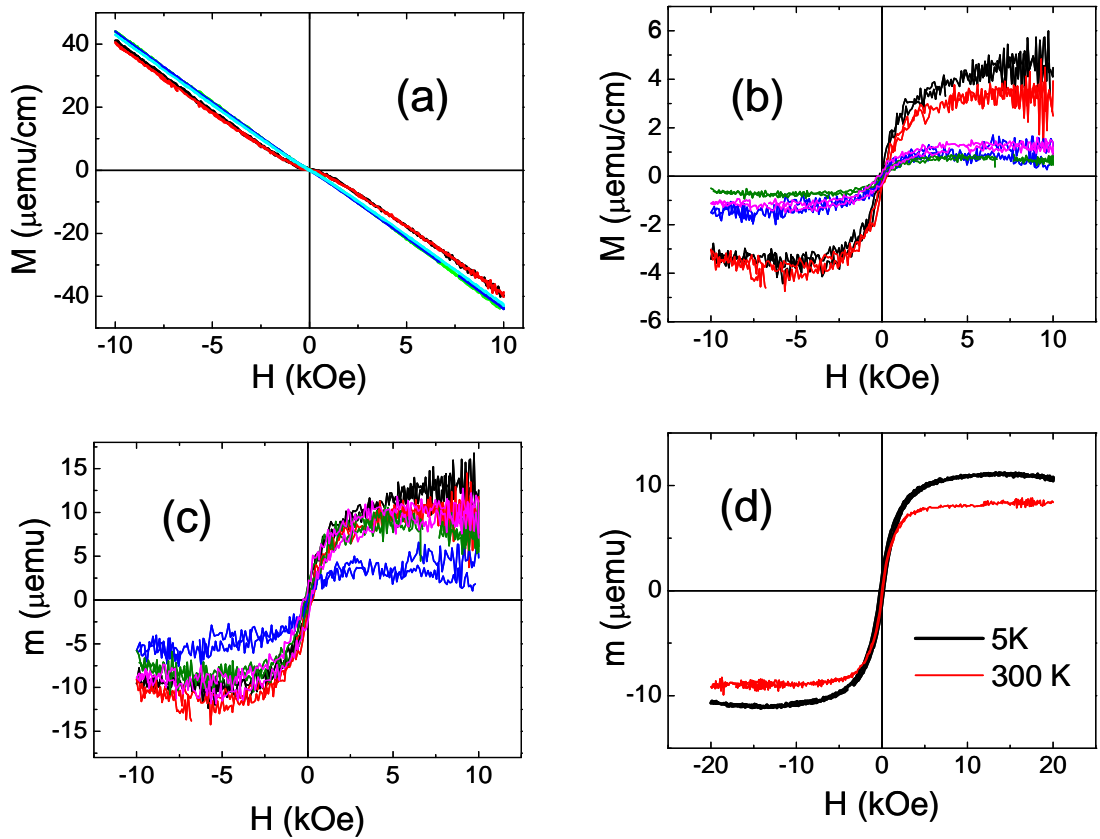

**Figure 1**. (a) Magnetization curves at 300 K from KT pieces of different lengths measured with a VSM; (b) the curves after subtracting a diamagnetic component; (c) the FM part of the curves (not normalized to the piece length); (d) FM part of the magnetization curve measured at 5 and 300 K.

Supporting the previous conclusion, the following experiment was performed: a piece of KT stored in air was measured and presented a FM signal of 5μemu. The piece was removed from the magnetometer and with the help of a clean cutter, 1 mm of the lateral parts of the piece was removed. After this operation, the piece exhibited again pure diamagnetic behavior. This experiment indicates that the borders of the KT, which are continuously exposed to the air and result sticky, are preferential positions for the deposition of dust particles while the center of the tape is more protected and can only be contaminated during the tape handling. Figure 2a shows the magnetization curves of a KT at 5 and 300 K. Differences in the diamagnetic susceptibility indicate the presence of a paramagnetic (PM) component; actually the thermal dependence of the magnetization (figure 2b) shows a Curie-like curve (in addition to a thermally independent diamagnetic background) which is 50% larger at 50 K than at 300 K. Differently to the FM signal, this PM susceptibility is found to be about the same for the different KT pieces and scales with the mass, indicating that it is not due to contamination during handling but to the presence of paramagnetic (probably iron) impurities in the tape. According to figure 2b this PM contribution can be neglected at RT within a resolution of $10^{-10}$ emu/Oe.

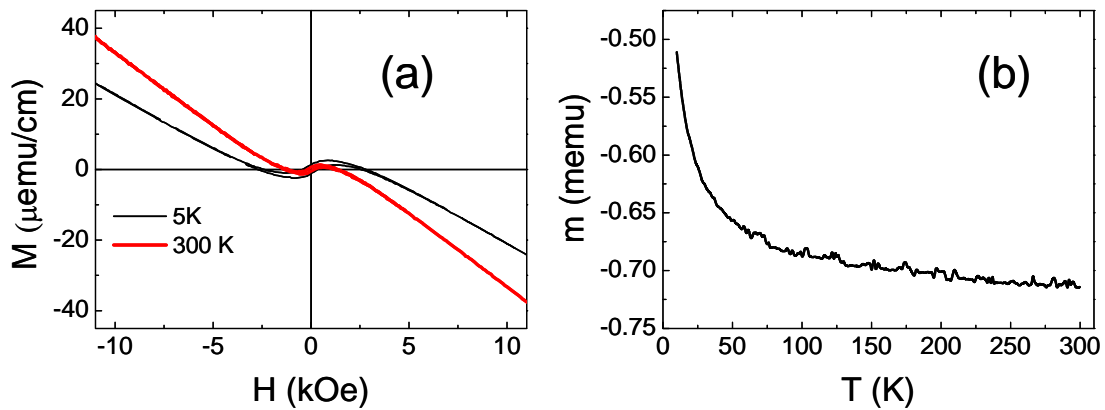

**Figure 2.** (a) Magnetization curves from a piece of KT at 5 and 300 K. (b) Thermal dependence of the magnetic moment under an applied field of 20 kOe.

Those experiments point out that FM and PM signals typically in the range $10^{-4}$ – $10^{-5}$ emu may appear when using KT. Cutting the edges or measuring the KT alone before or after the sample measurements can be useful also to identify this contamination. However, it is important to remark that any handling of the tape can increase the deposition of the dust particles (enhancing the FM signal) and when removing the tape from any surface some particles may remain stuck on the substrate (reducing the FM signal).

## 2.2. Iron tools

Contamination of a sample when manipulated with iron tools consists on the deposition of small Fe microparticles in the contact point between the tool and the sample. The problem is particularly severe for oxides due to their hardness. Covalent bonds present in oxides are much stronger than the metallic bonds so when a metallic tool comes in contact with an oxide, usually some of the metal particles can be removed from the tool and incorporated to the oxide.

Many oxides have been claimed to be FM at the nanoscale (actually some recent publications claim that all the oxides are FM at the nanoscale [15] so, when analyzing the magnetic properties of this nano-oxides, it is particularly important to ensure that there is no contamination due to the use of iron based tools in any step of the sample handling as previously reported [21].

Figure 3 shows the magnetization curve at room temperature of a glass (silicon oxide) substrate 3x3x0.5 millimeters before and after pressing one of the borders with steel tweezers. As it is observed, the contact with the iron tool produces a FM signal of 0.1 memu.

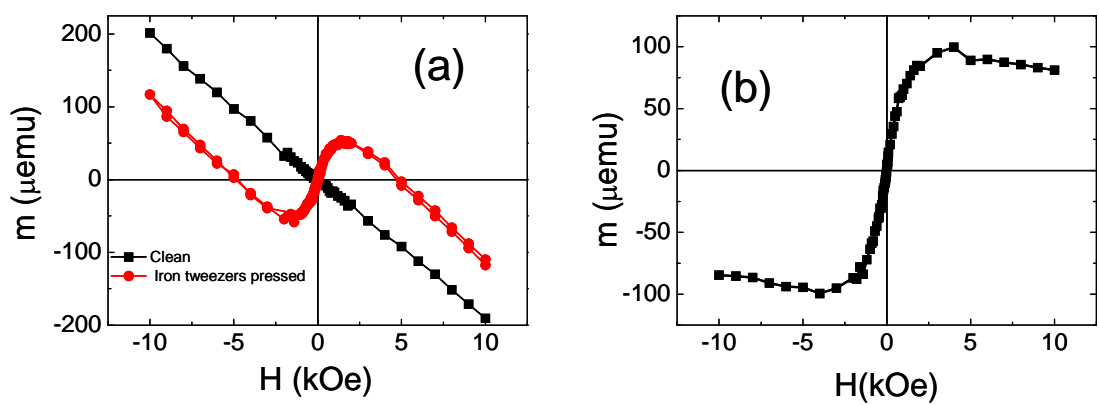

**Figure 3.** (a) Magnetization curves from a clean silica glass (3x3x0.5 mm and after pressing the border with iron tweezers at 300 K. (b) Difference between both curves.

When oxides are in contact with metallic elements, it is not necessary macroscopic strong forces to remove some metallic particles from the tool and incorporate them to the oxide. Even very weak macroscopic forces can produce high stress in the small contact region of the oxide and the tool. As an example, figure 4 shows magnetization curves at 300 K from alumina powder after sieving with inox steel and nylon sieves. After using the steel tool, the sample exhibits a FM signal (superimposed to the diamagnetic one) with a saturation magnetization $M_S$ ~0.5 memu/g which is not present when using a non- metallic (nylon) sieve. This FM signal scales with the sample mass demonstrating that this is a homogeneous contamination. The micrograph of the powder after sieving with inox steel shows the large alumina particles and some small (submicrometer) ones which according to EDS are iron rich particles from the sieve. This effect was also measured in other oxides, including ZnO.

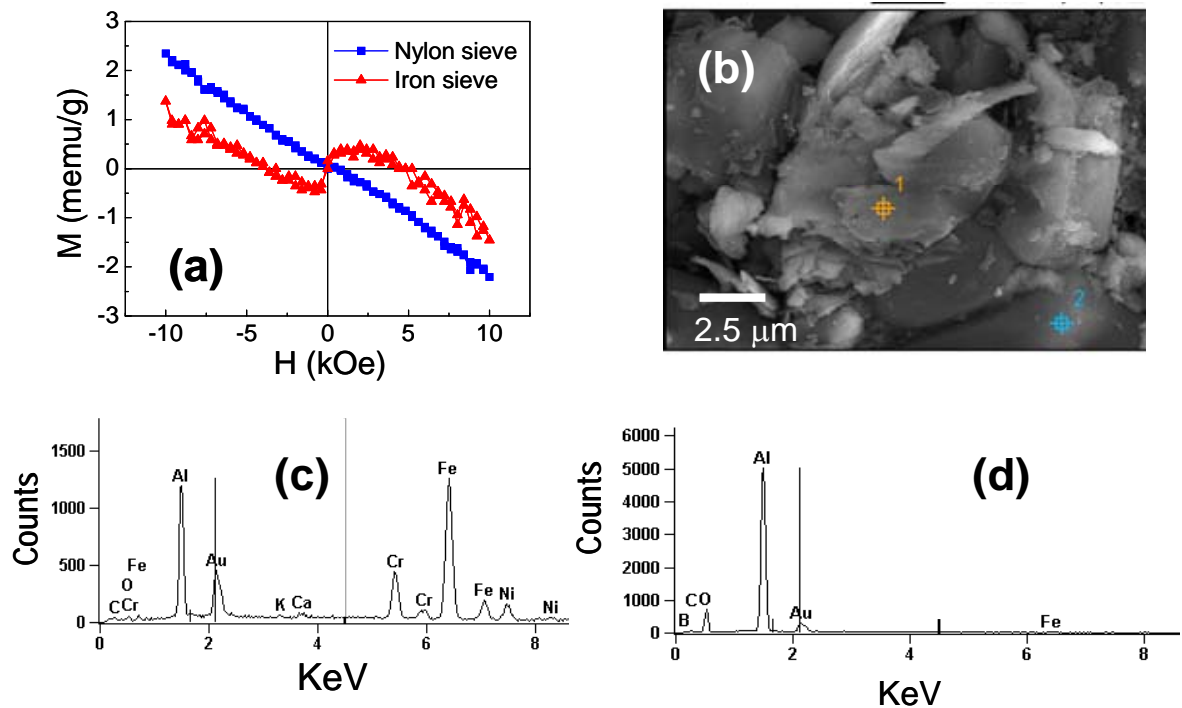

**Figure 4.** (a) Magnetization curves of $Al_2O_3$ passed through nylon and steel at 300K. (b) Micrograph from the samples sieved with the steel tool. EDS analysis corresponding to points (1) and (2) are presented in (c) and (d) respectively.

## 2.3. Gelatin capsules, cotton and plastic straws.

Gelatin capsules are commonly used to hold powders samples that are later pressed with cotton to avoid movements during magnetic measurements (specially in the case of VSM magnetometers). The mass of those gelatin capsules is usually 30-50 mg. We measured the magnetization curves of 5 gelatin capsules and they exhibited diamagnetic behavior without any deviation up to a sensitivity of $10^{-6}$ emu (that is, $3 \cdot 10^{-5}$ emu/g).

Commercial cotton is a soft, staple fiber that grows around the seeds of the cotton plant. Nominally, it is a diamagnetic material with susceptibility that ranges between $10^{-6}$ and $10^{-7}$ emu/g·Oe depending on the supplier. The cotton used in labs is usually commercial cotton for general purposes so no special procedures to avoid contamination are considered during its production. Typical mass of cotton used to fill capsules with powder samples is ~ 50 mg. As figure 5 shows, commercial cotton exhibits the expected diamagnetic behavior (with susceptibility of $10^{-6}$ emu/g·Oe) plus an additional anhysteretic ferromagnetic-like contribution of ~5 memu ($M_S$~ 0.1 emu/g). The shape of this FM contribution matches to that found in KT and samples handled with iron tools (see previous sections) suggesting that it is also due to the presence of small Fe (or Fe containing particles).

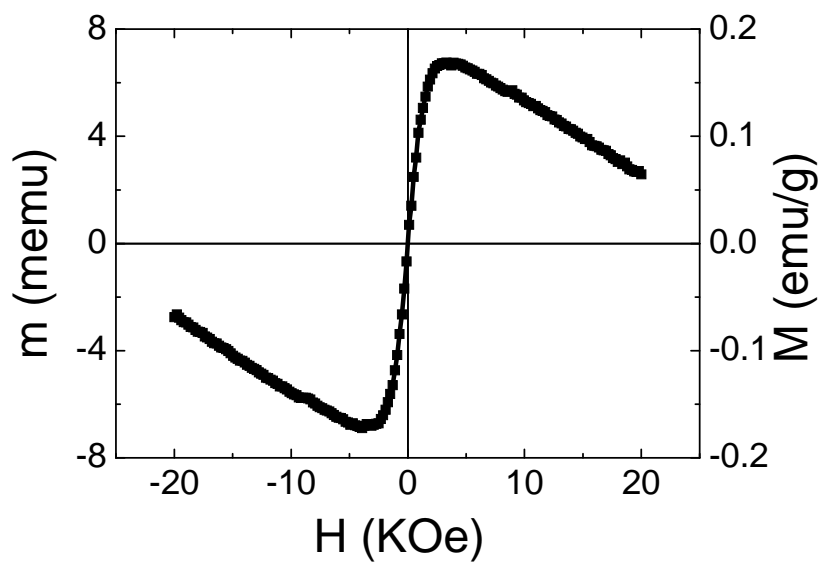

**Figure 5.** Magnetization curve from a piece of 50 mg of commercial cotton at 300K..

Plastic straws are used to place samples in SQUID sampleholders. Usually, no magnetic signals are detected from plastics straws in the SQUID. It is noteworthy that a homogeneous distribution of magnetic impurities along the straw will provide no magnetic signal as they are large enough so the net magnetic flux across the coils does not change when the straw moves through the coil. However, any asymmetry or inhomogeneity in the straw magnetic profile will give rise to a magnetic signal.

Figure 6 shows magnetization curves at 5 K from a silicon substrate placed on plastic straw without deformation and the same sample placed onto another straw that was intentionally pressed to induce a plastic deformation at the sample position. As it can be observed, after deformation, a weak magnetic signal of the order of $10^{-6}$ emu is detected that could be ascribed to the deformation of the straw; if impurities are present in the straw, deformation can modify locally the flux lines arising a magnetic signal.

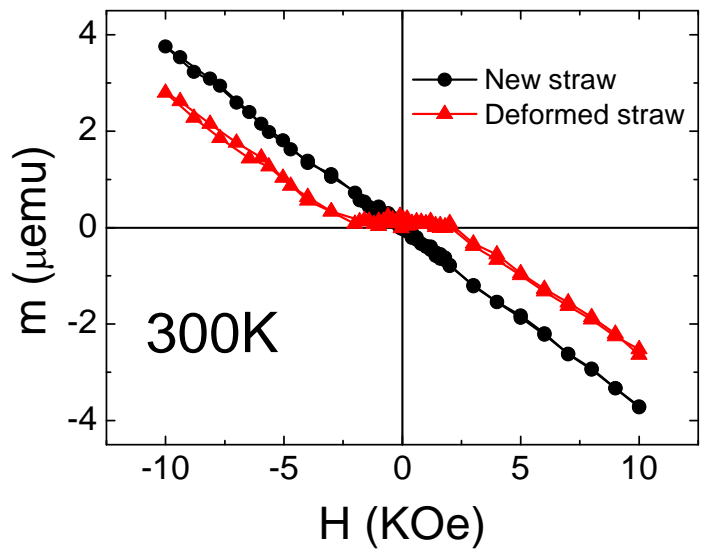

**Figure 6.** Magnetization curves at 300 K from a Si substrate measured on a non deformed (black circles) and a plastically deformed (red triangles) straw measured with a SQUID.

2.4 Inks and Silver paint

Inks are commonly used to mark samples in order to identify them or determine their orientation. Many commercial inks include magnetic impurities, specially those corresponding to red and red-based colors (pink, magenta, etc..) as the red color is achieved mainly by iron oxide. We found magnetic signals in silicon substrates up to $4\cdot10^{-5}$ emu when painting a cross (2mm×2mm) on it with a red pencil. Half of the value was found for blue pencil from the same supplier. The shape of the FM signal was similar to that presented in figure 1 corresponding to KT.

Silver paint is used to place electrical contact on samples, mainly for transport measurements. Although nominally it does not contain iron, we experimentally found (figure 7) a magnetic signal in silver paint that scales with the mass, indicating an intrinsic magnetic contamination of this material. The value of the magnetic signal is of the order of ~1 emu/g of dry silver paint (that is, measuring the mass after the paint is dried) which obviously will change depending on the silver paint features, composition and supplier. However, it turns out that a signal of $10^{-5}$ emu could be arisen by 100 µg of silver paint, not easy to detect by eye inspection. Therefore, if silver paint is used to electrically contact a sample, it is highly recommended to perform the magnetic characterization before the contacts are done; if it is not possible, it is important to characterize the silver paint used and analyze its possible contribution to the measured magnetic moment.

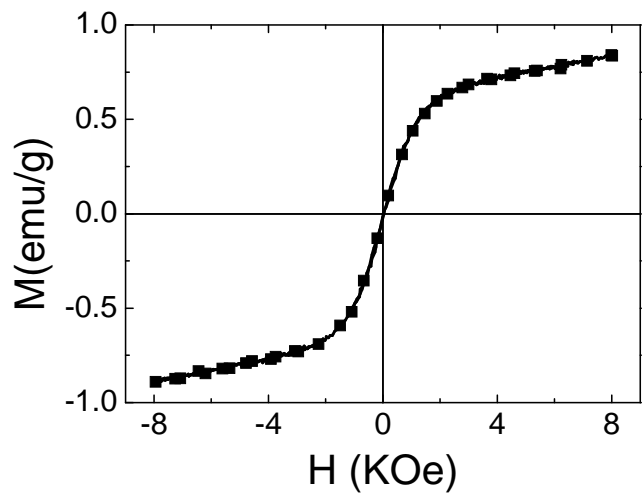

**Figure 7**. Magnetization curve at 300 K from dried silver paint.

2.5 Anisotropy artifacts

A relatively common feature of the new surface magnetism is the different value of $M_S$ value when the magnetic field is applied in parallel and perpendicular to the sample surface [3,4,23]. Most of the magnetic measurements techniques (VSM, SQUID; extraction techniques) are based on the induction on a coil when the magnetic flux across the coil is modified. As figure 8 illustrates, the magnetic flux across a coil does not depend only on the value of the magnetic moment but also on its position respect to the coil center (actually, most magnetometers user manuals indicate that the calibration of the equipment is only exact for samples with the same shape than the standard). The dependence of the magnetic signal on the sample positioning has been extensively studied [24,25].

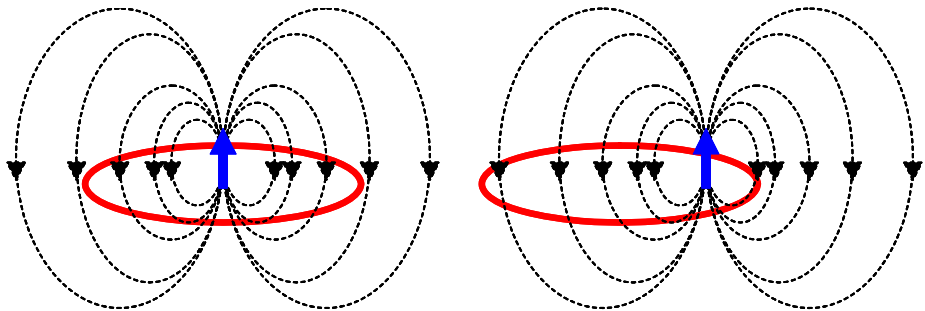

**Figure 8.** Illustration of the magnetic flux across a coil created by a punctual magnetic moment. The net flux is the magnetic moment minus the number of field lines that closes inside the coil; this latter term depends on the position of the magnetic moment respect to the center of the coil (radial symmetry).

Consider an iron impurity attached to a sample with a magnetic moment $10^{-5}$ emu as those due to contamination by the use of iron tools. Figure 9 presents the magnetic flux across the coil from a magnetic moment in the coil plane and oriented perpendicular to the coil plane as a function of the distance to the center of the coil.

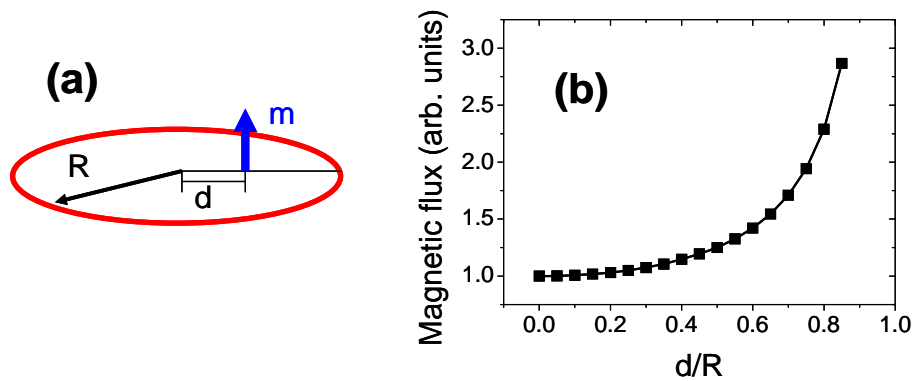

**Figure 9**. Calculated magnetic flux across a coil of radio R due to magnetic moment m located in the coil plane at a distance d to the center of the coil.

Therefore, the magnetic signal due to inhomogeneous contamination will depend on its position. For samples contaminated with small Fe-based microparticles, measurements with different orientation can lead to different values of $M_S$ because of the different position of the contamination particles and not related to any magnetic anisotropy. Simply turning the sample to measure in plane and perpendicular orientation can modify the position of the contamination particle leading to different magnetization curves. As an example, figure 10 presents the magnetization curves at room temperature of a silicon oxide substrate (3x3x0.5 mm) intentionally contaminated by pressing with iron tweezers on the borders (the place where the samples are normally grab while handling) measured applying the field parallel and perpendicular to the plane in a SQUID. While the diamagnetic contribution (from the silica substrate) is about the same, the ferromagnetic contribution is 40% larger when measuring in perpendicular. In this later situation, the contamination area will be closer to the border of the coil, arising a larger signal in agreement with the data in figure 9.

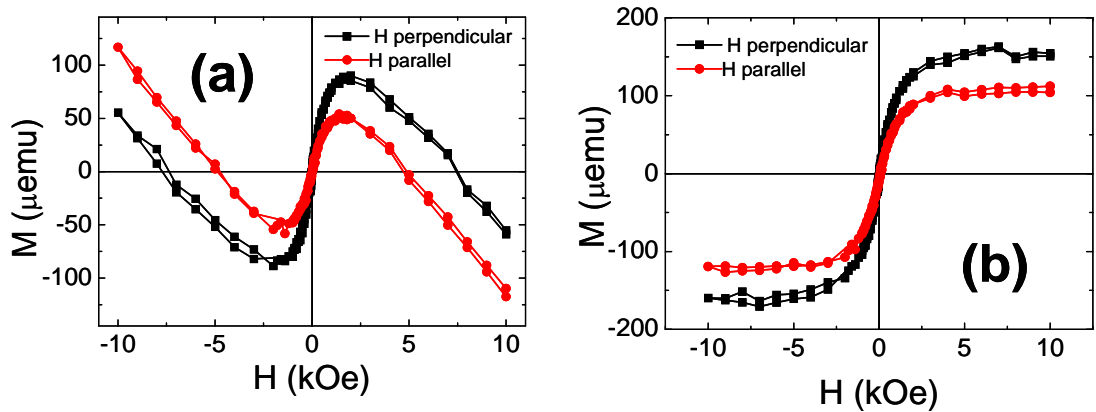

**Figure 10**. (a) Magnetization curves at 300 K measured with a SQUID from a silica glass (3x3x0.5 mm) after pressing the border with iron tweezers and placing the sample in plane and perpendicular to the magnetic field direction. (b) The same curves after subtracting the diamagnetic component.

## 3. Conclusions

In summary, we showed here that standard procedures used when handling samples for magnetic measurements may induce the appearance of magnetic signals of the order of $10^{-4}$ emu. There are undoubtedly some other signal arising from different sources than those identified here. These signals are negligible when measuring large quantities of magnetic materials with high $M_S$, but must be considered when measuring very weak signal from materials with a small number of magnetic atoms and total magnetic moments of the order of $10^{-4}$ emu or below. There are mainly two kind of spurious signals:

a)  Those due to homogeneous sample contamination which are proportional to the samples mass. They can be discarded by chemical analysis providing impurity levels below those required to justify the measured signals.

b)  Contamination related to the measurement procedures that introduces a signal irrespective of the sample mass. This kind of contamination must be considered before mass normalization (i.e., considering the total value of emu measured by the magnetometer) and check if such magnetic moment could be arised by any other source than the sample. These signals are particularly tricky as, if they are not properly corrected, when normalizing to the sample mass, magnetization values may become huge.

Although we presented here some guidelines to identify these effects, the spurious signals are largely dependent on the particular experimental set-up, materials and measurement protocols used at each lab. Thus, it is not possible to establish standard procedures to avoid them but each lab should carry on their own tests to determine the reliability of their magnetic measurements.

## 4. Acknowledgements

M. Multigner and L. Pérez are acknowledged for valuable suggestions and advice. This work has been supported by the EU through the project BONSAI (LSHB-CT-2006-037639, www.bonsai-project.eu) and by the Spanish Council for Scientific Research through the projects MAGIN -CSIC 2006-50F0122.